\documentclass[12pt]{iopart}
\usepackage{amsfonts}

\renewenvironment{thebibliography}[3]
    {\frenchspacing
     \normalsize
     \baselineskip=12pt
     \begin{list}{$^{\arabic{enumi}}$}
        {\usecounter{enumi}\setlength{\parsep}{0pt}
     \setlength{\leftmargin 17pt}{\rightmargin 0pt}   %FOR 10--99 ITEMS
         \setlength{\itemsep}{6pt} \settowidth
    {\labelwidth}{#1.}\sloppy}}{\end{list}}
\makeatletter
\renewcommand{\@cite}[2]{$\scriptsize ^{#1\if@tempswa , #2\fi}$}
\makeatother

\newcommand{\om}{\omega}
\newcommand{\Om}{\Omega}
\newcommand{\dl}{\delta}
\newcommand{\bk}{{\bf k}}
\newcommand{\ve}{\varepsilon}
\newcommand{\ep}{\varepsilon}
\newtheorem{theorem}{Theorem}
\newtheorem{lemma}{Lemma}
\newtheorem{definition}{Definition}
\newtheorem{remark}{Remark}
\newtheorem{example}{Example}

\begin{document}

\title[]{The multi-time correlation functions, free white noise,
and the generalized Poisson statistics in the low density
limit}

\author{A. N. Pechen}

\address{Princeton University, Department of  Chemistry,
         Frick Laboratory, Princeton, NJ 08544-1009}

\ead{\mailto{apechen@princeton.edu}}

\begin{abstract}
In the present paper the low density limit of the
non-chronological multitime correlation functions of boson
number type operators is investigated. We prove that the
limiting truncated non-chronological correlation can be
computed using only a sub-class of diagrams associated to
non-crossing pair partitions and thus coincide with the
non-truncated correlation functions of suitable free
number operators. The independent in the limit subalgebras
are found and the limiting statistics is investigated. In
particular, it is found that the cumulants of certain
elements coincide in the limit with the cumulants of the
Poisson distribution. An explicit representation of the
limiting correlation functions and thus of the limiting
algebra is constructed in a special case through suitably
defined quantum white noise operators.
\end{abstract}

\newpage
\section{INTRODUCTION}
The reduced dynamics of a quantum open system interacting
with a reservoir in certain physical regimes is
approximated by Markovian master equations. These regimes
include the weak system--reservoir interactions and dilute
reservoirs and in the theoretical framework they are
described by certain limits. For a weakly interacting
system one considers the limit as the coupling constant
goes to zero (Weak Coupling Limit, WCL) whereas for a
dilute reservoir one considers the limit as the density of
the reservoir goes to zero (Low Density Limit, LDL) and an
appropriate time rescaling should be performed in order to
get a non-trivial limit. The Markovian reduced dynamics in
these limits is considered in the review papers by Spohn
and Lebowitz\cite{sl}$^,$\cite{spohn}. The reduced
dynamics in the LDL was considered in details later by
D\"umcke\cite{d} using the method based on the quantum
Bogoliubov--Born--Green--Kirkwood--Yvon hierarchy.

The total dynamics in these limits is governed by various
quantum stochastic equations. There is a unique up to now
approach, called the {\it stochastic limit method}, which
allows an efficient derivation of the stochastic equations
in the WCL. This approach is based on the quantum white
noise technique and was developed by Accardi, Lu, and
Volovich\cite{alv}.

The convergence of the evolution operator of the total
system in the LDL to a solution of a quantum stochastic
equation was proved by Accardi and Lu\cite{al} and by
Rudnicki, Alicki, and Sadowski\cite{ras}. Recently the low
density limit was investigated with the quantum white
noise technique\cite{apv1}$^,$\cite{apv2}. This technique,
well developed for the WCL, was non-trivially modified to
include the LDL and for this case was called {\it the
stochastic golden rule for the low density limit}. This
technique was applied to the derivation of the quantum
stochastic equations in the LDL. An advantage of the
obtained equations is that they, in contrast with the
exact Schr\"odinger equation, are explicitly solvable. At
the same time they provide a good approximation of the
exact dynamics.

The approach of\cite{apv1}$^,$\cite{apv2} uses the so
called Fock-antiFock representation for the canonical
commutation relations (CCR) algebra (this representation
is unitary equivalent to the Gel'fand--Naimark--Segal
representation). The difficulty with this approach is that
the creation and annihilation operators in the
Fock-antiFock Hilbert space do not describe creation and
annihilation of physical particles and thus do not have
direct physical meaning. To avoid this difficulty the
investigation of the LDL directly in terms of the physical
fields was performed\cite{p}. Using this approach the
chronological correlation functions in the LDL were found
and the corresponding stochastic equations derived.

In the present paper we investigate the low density limit
of the non-chronologically ordered correlation functions
of boson number type operators. The investigation is
related with {\it ab initio} derivations of quantum
stochastic equations describing quantum dynamics of a test
particle interacting with a dilute gas. We find the
limiting truncated correlation functions of the number
type operators and show that they can be computed by
representing the number operators through creation and
annihilation operators and then considering only a
sub-class of diagrams associated to non-crossing pair
partitions. This fact allows to represent the limiting
truncated correlation functions as the non-truncated
correlation functions of number operators of a free
quantum white noise thus making a connection with the
Voiculescu free probability theory. We find the limiting
statistics and show that the cumulants of certain elements
coincide in the limit with the cumulants of the Poisson
distribution.

The free probability theory was developed by Voiculescu
around 1985 as a way to deal with von Neumann algebras of
free groups. Then the theory was separated from this
special context and began to develop as an independent
field. In particular, applications of the free
independence theory to random matrices were found. The
details of free probability theory and its applications to
random matrices could be found, for example, in
references\cite{vdn}$^,$\cite{v}.

Expectations of free random variables are characterized by
diagrams associated to non-crossing pair partitions. The
vanishing of crossing diagrams in the stochastic weak
coupling limit for nonrelativistic QED and for the
Anderson model was found in Refs\cite{alv} and\cite{alm},
respectively, thus making a connection between the WCL and
free probability. The WCL is typically described by the
quantum Boltzmann statistics\cite{alv}. In Ref\cite{aav} a
generalized version of Boltzmann commutation relations,
the so called entangled commutation relations, was found
in the weak coupling limit for nonlinear interactions and
possible applications to photon splitting cascades were
discussed.

The investigation of the multitime non-chronologically
ordered correlation functions could have a connection with
the behavior of fluctuations in certain asymptotic
regimes. The latter is described in the review paper by
Andries, Benatti, De Cock and Fannes\cite{abcf}. In that
approach the limiting statistics is defined in terms of
ground state distribution determined by non trivial pair
partitions. The authors conjecture the appearance of
exotic statistics in certain asymptotic regimes. The
asymptotic fluctuations are the limiting correlation
functions of appropriate centered elements and thus the
results of the present paper could be applied to study the
fluctuations in the low density limit.

In Sec.~II the truncated non-chronologically ordered
correlation functions are defined and their low density
limit is established (Theorem~\ref{mainth}). In Sec.~III
the irreducible diagrams (pair partitions) which
contribute to the limiting correlation functions are found
(Theorem~\ref{th1}). In Sec.~IV the limiting truncated
correlation functions are represented as correlation
functions of a suitable free white noise. In Sec.~V we
identify the independent in the limit subalgebras
(Theorem~\ref{th2}) and calculate the limiting cumulants
which for some elements coincide with the cumulants of the
Poisson distribution (Theorem~\ref{th4}). In Sec.~VI an
explicit representation of the limiting correlation
functions and thus of the limiting algebra is constructed
for a special case by using suitable quantum white noise
operators.

\section{THE CORRELATION FUNCTIONS IN THE LDL}
We begin this section with construction of a general class
of non-commutative probability spaces relevant for the
investigation of the low density limit. The framework of a
$*$-probability space is used. A relation between the
objects defined in this section and the model of a test
particle interacting with a dilute gas is given in
Appendix A.

\begin{definition}
A {\bf $*$-probability space} is a pair $({\cal A},\om)$, where
${\cal A}$ is a unital $*$-algebra over $\mathbb C$ and  $\om:{\cal
A}\to\mathbb C$ is a state, i.e., a linear normalized, $\om(1_{\cal
A})=1$, and strictly positive functional.
\end{definition}

Let ${\cal H}$ be a Hilbert space with inner product
denoted by $\langle\cdot,\cdot\rangle$ (called as one
particle Hilbert space), $\{S_t\}_{t\in\mathbb R}$ a one
parameter unitary group in $\cal H$ (a one particle free
evolution), $\hat n$ a bounded positive operator in ${\cal
H}$ (density operator) such that $\forall t\in\mathbb R$,
$S_{-t}\hat nS_t=\hat n$, and $B$ a countable set of real
numbers.

Let $\Gamma({\cal H})$ be the symmetric Fock space over
$\cal H$. For any trace class self-adjoint operator $T$
acting in $\cal H$ we denote by $N(T)\equiv\rmd\Gamma(T)$
its second quantization operator in $\Gamma({\cal H})$ and
extend this definition by complex linearity to the set of
all trace class operators ${\cal T}({\cal H})$. For any
$T\in{\cal T}({\cal H})$, $\om\in B$, and a positive
number $\ep>0$ we define the following operator in
$\Gamma({\cal H})$:
\begin{equation}\label{n0}
N_{T,\om,\ve}(t):=\frac{\rme^{-\rmi
t\om/\ve}}{\ve}N(S_{t/\ve}TS_{-t/\ve})
\end{equation}
Let $L(\mathbb R)=\bigcap_{p\in\mathbb N} L^p(\mathbb R)$,
where $L^p(\mathbb R)$ is the space of $p$-power
intergable functions over $\mathbb R$. For any open subset
$\Lambda\subseteq\mathbb R$ let $L(\Lambda)$ be the set of
functions from $L(\mathbb R)$ with support in $\Lambda$.
We denote by ${\cal A}_{\Lambda,\ep}$ the $*$-algebra
generated by operators $N_{T,\om,\ve}(\varphi):= \int\rmd
t\varphi(t) N_{T,\om,\ve}(t)$ with $T\in {\cal T}({\cal
H})$, $\om\in B$, $\varphi\in S(\Lambda)$ and denote
${\cal A}_\ep:={\cal A}_{\mathbb R,\ep}$.

Let $A^\pm(g)$, $g\in{\cal H}$ be the creation and
annihilation operators in $\Gamma({\cal H})$ [we denote in
the sequel $A^-(g)\equiv A(g)$] with the canonical
commutation relations $[A(f),A^+(g)]=\langle f,g\rangle$
and let ${\cal A}_{\rm CCR}$ be the algebra of polynomials
in $A^\pm(\cdot)$. Any operator $N(T)$ can be represented
in terms of the creation and annihilation operators. For
example, if $T=|f\rangle\langle g|$, where we use Dirac's
notations for elements $f,g\in{\cal H}$, then
$N(T)=A^+(f)A(g)$. An arbitrary operator $N(T)$ can be
expressed in terms of $A^\pm$ using the fact that any
trace class operator $T$ is a limit of finite rank
operators. Thus the algebra ${\cal A}_\ep$ is a subalgebra
of ${\cal A}_{\rm CCR}$.

Let $\om_{\hat n}$ be a gaussian gauge-invariant mean-zero
state on ${\cal A}_{\rm CCR}$ with the two point
correlation function $\om_{\hat n}(A^+(f)A(g)):=\langle
g,\hat nf\rangle$ (thus $\om_{\hat n}(N(T))=\Tr (\hat n
T)$ and here we use the assumption for $T$ being trace
class). Denoting by the same symbol its restriction to
${\cal A}_{\Lambda,\ep}$, we finally have for any $\ep>0$
and for any open subset $\Lambda\subseteq\mathbb R$ the
$*$-probability space $({\cal
A}_{\Lambda,\ep},\om_{\ep\hat n})$.

\begin{remark} The condition $\forall t$: $S_{-t}\hat nS_t=\hat n$
leads to the invariance of the state $\om_{\hat n}$ under
the free evolution generated by $S_t$.
\end{remark}

With the notations above we define the non-chronologically
ordered multitime correlation functions as
\begin{eqnarray}\label{w0}
W_{\ve,\hat n,T_1,\om_1,\dots, T_n,\om_n}(t_1,\dots,
t_n)&:=&\om_{\ve\hat n}(N_{T_1,\om_1,\ve}(t_1)\dots
N_{T_n,\om_n,\ve}(t_n))\\
W_{\ve,\hat n,T_1,\om_1,\dots, T_n,\om_n}(\varphi_1,\dots,
\varphi_n)&:=&\om_{\ve\hat
n}(N_{T_1,\om_1,\ve}(\varphi_1)\dots
N_{T_n,\om_n,\ve}(\varphi_n))\label{w1}
\end{eqnarray}
We will use for the correlation functions~(\ref{w0})
and~(\ref{w1}) also the shorter notations
$W_\ve(t_1,\dots, t_n)$ and $W_\ve(\varphi_1,\dots,
\varphi_n)$. The reason for introducing the averaged
operators $N_{T,\om,\ep}(\varphi)$ and the averaged
correlation functions~(\ref{w1}) is that, as we will show
below, the non-averaged operators $N_{T,\om,\ep}(t)$ and
the correlation functions~(\ref{w0}) in the limit as
$\ep\to 0$ become singular distributions. Clearly, one has
the relation
\[
W_\ve(\varphi_1,\dots,\varphi_n)=\int\rmd t_1\dots\rmd t_n
W_\ve(t_1,\dots, t_n)\varphi_1(t_1)\dots\varphi_n(t_n)
\]

\begin{definition}\label{truncated}
The truncated correlation functions $W^T_\ve(t_1,\dots,
t_n)$ are defined for $n=1$ by $W^T_\ve(t_1):=W_\ve(t_1)$
and for $n>1$ by induction through the relation:
\begin{eqnarray*}
W_\ve(t_1,\dots, t_n)&=&W^T_\ve(t_1,\dots,
t_n)+\sum\limits_{l=2}^n{\sum}'
W^T_\ve(t_{i_1},\dots, t_{i_{k_1}})\\
&&\times W^T_\ve(t_{i_{k_1+1}},\dots, t_{i_{k_2}})\dots
W^T_\ve(t_{i_{k_l}},\dots, t_{i_n})
\end{eqnarray*}
where $\sum'$ is the sum over $i_1<i_2<\dots<i_{k_1}$,
$i_{k_1+1}<\dots<i_{k_2},\dots,$ $i_{k_l+1}<\dots<i_{n}$.
\end{definition}

The truncated correlation functions are often used in
quantum field theory and in quantum kinetic
theory\cite{b}. They entirely determine the corresponding
non-chronological correlation functions. Thus the
investigation of the limit of the non-chronological
correlation functions can be reduced to the investigation
of the limit of the truncated correlation functions.

We define the "projection" $P_E:=(2\pi)^{-1}\int\rmd
tS_t\rme^{-\rmi tE}$ [it has the property
$P_EP_{E'}=\dl(E-E')P_E$] and for any $k=1,2,\dots,n$
denote $\tilde\om_k=\om_n+\dots+\om_k$. The following
theorem states the low density limit of the truncated
correlation functions.

\begin{theorem}\label{mainth}
One has the limit in the sense of distributions in variables
$t_1,\dots, t_n$:
\begin{eqnarray}
\fl\lim\limits_{\ve\to 0}W^T_{\ve,\hat
n,T_1,\om_1,\dots,T_n,\om_n}(t_1,\dots, t_n)
&=&(2\pi)^{n-1}\dl(t_2-t_1)\dots\dl(t_n-t_{n-1})\nonumber\\
&&\times\dl_{\tilde\om_1,0}\int\rmd E\Tr\Bigl[\hat n
P_{E+\tilde\om_1}T_1P_{E+\tilde\om_2}T_2\dots
P_{E+\tilde\om_n}T_n\Bigr]\label{eq5}
\end{eqnarray}
where $\Tr$ denotes trace and $\dl_{\tilde\om_1,0}$ is the
Kronecker delta symbol.
\end{theorem}
The theorem is a corollary of Theorem~\ref{th1} from
Section~\ref{sec1}.

\section{THE NON-TRIVIAL DIAGRAMS}\label{sec1}
In the present section we investigate the low density
limit of the non-chronologically ordered correlation
functions for the particular case of operators of the form
$T_l=|f_l\rangle\langle g_l|$ and find the diagrams which
are non-trivial in the low density limit.

In order to simplify the notations we will use the
following {\bf energy representation} for the creation and
annihilation operators:
\[
A^+_l:=\frac{\rme^{\rmi t_lE_l/\ve}}{\sqrt{\ve}}A^+(P_{E_l}f_l);
\qquad A_l:=\frac{1}{\sqrt{\ve}}A(S_{t_l/\ve}g_l)
\]
(a slightly different version of the energy representation
was introduced in\cite{apv1}). One has
$N_{T_l,\om_l,\ve}(t_l)=\rme^{-\rmi t_l\om_l/\ep}\int \rmd
E_lA^+_lA_l$. Notice that the operator $A^+_l$ is not the
adjoint of $A_l$. The symbols $A_l$, $A^+_l$ are used only
to simplify the notations below.

A multitime correlation function can be expressed using
Gaussianity of the state $\om_{\hat n}$ and the energy
representation for the creation and annihilation operators
as
\begin{eqnarray}\fl
W_{\ve,\hat n,T_1,\om_1,\dots, T_n,\om_n}(t_1,\dots,
t_n)&=& \exp\Bigl(-\rmi\sum\limits_{l=1}^n \om_l
t_l/\ep\Bigr) {\sum}'\int\rmd E_1\dots\rmd E_n\om_{\ve\hat
n}(A^+_{i_1}A_{j_1})\dots\nonumber\\
&&\times \om_{\ve\hat n}(A^+_{i_k}A_{j_k})\om_{\ve\hat n}(A_{j_{k+1}}A^+_{i_{k+1}})\dots
\om_{\ve\hat n}(A_{j_n}A^+_{i_n})\label{eq1}
\end{eqnarray}
where $\sum'$ is the sum over $k=1,\dots, n$,
$1=i_1<i_2<\dots<i_k$, $j_{k+1}<\dots<j_n$, $i_l\le j_l$
for $l=1,\dots,k$ and $j_l<i_l$ for $l=k+1,\dots,n$. The
sum contains terms of the form
\begin{equation}\label{10}
\om_{\ve\hat n}(A^+_{i_1}A_{j_1})\dots \om_{\ve\hat
n}(A^+_{i_k}A_{j_k}) \om_{\ve\hat n}(A_{j_{k+1}}A^+_{i_{k+1}})\dots
\om_{\ve\hat n}(A_{j_n}A^+_{i_n})
\end{equation}
To each such term we associate a diagram by
pairing in the string $A^+_1A_1A^+_2A_2\dots A^+_nA_n$ the operators
$A^+_{i_l}$ and $A_{j_l}$ for $l=1,2,\dots n$.

\begin{definition}
We say that the expression~(\ref{10}) corresponds to a reducible
diagram if there exists a nonempty subset $I\subset\{1,\dots, n\}$
(strict inclusion) such that $i_l\in I\Leftrightarrow j_l\in I$.
Otherwise we say that the expression~(\ref{10}) corresponds to an
irreducible diagram.
\end{definition}

An important property of the truncated correlation
functions (Def.~\ref{truncated}) is that they keep only
all irreducible diagrams. The following are the examples
of irreducible (first) and reducible (second) diagrams for
$n=2$:\\
\noindent
\begin{picture}(50,0)
 \put(78,5){\line(1,0){46}}\put(78,5){\line(0,-1){13}}\put(124,5){\line(0,-1){13}}
 \put(94,0){\line(1,0){14}}\put(94,0){\line(0,-1){8}}\put(108,0){\line(0,-1){8}}
 \put(188,5){\line(1,0){16}}\put(188,5){\line(0,-1){13}}\put(204,5){\line(0,-1){13}}
 \put(217,5){\line(1,0){16}}\put(217,5){\line(0,-1){13}}\put(233,5){\line(0,-1){13}}
\end{picture}
\begin{equation}\label{eq01}
A^+_1A_1A^+_2A_2\qquad\qquad A^+_1A_1A^+_2A_2
\end{equation}
Given an reducible diagram, one can represent the set
$\{1,\dots n\}$ as a union of several disjoint subsets
$I_1,\dots, I_l$ such that the diagram contains only
pairings between operators with indices from the same
subsets. In this sense a general reducible diagram can be
represented as a union of mutually disjoint irreducible
diagrams. Examples of the truncated correlation functions,
the corresponding irreducible diagrams, and their limits
as $\ep\to 0$ for $n=1,2,3$ are given below.
\begin{example} \rm $n=1$. The invariance of the state under
the free evolution leads to the identity $W^T_\ve(t)\equiv
W_\ve(t)\equiv W_\ve(0)=\langle g_1,\hat n f_1\rangle$.
\end{example}
\begin{example}\rm $n=2$. One has
\begin{eqnarray}
W^T_\ve(t_1,t_2)&=&W_\ve(t_1,t_2)-W_\ve(t_1)W_\ve(t_2)=\int\rmd
E_1\rmd E_2\om_{\ve\hat
n}(A^+_1A_2)\om_{\ve\hat n}(A_1A^+_2)\nonumber\\
&=&\int\rmd E_1\rmd
E_2\frac{\rme^{\rmi(t_2-t_1)(E_2-E_1)/\ve}}{\ve}\langle
g_2, P_{E_1}\hat nf_1\rangle\langle g_1,(1+\ve \hat
n)P_{E_2}f_2\rangle\label{eq2}
\end{eqnarray}
This expression corresponds to the first (irreducible)
diagram in~(\ref{eq01}) which is non-zero in the limit.
Application of Lemma~\ref{lemma} (see Appendix B) to the
r.h.s. of~(\ref{eq2}) gives
\[
\lim\limits_{\ve\to 0}W^T_\ve(t_1,t_2)=2\pi\dl(t_2-t_1)\int \rmd
E\langle g_2, P_E\hat nf_1\rangle\langle g_1,P_Ef_2\rangle
\]
\end{example}
\begin{example}\rm $n=3$. One has
\begin{eqnarray*}
W^T_\ve(t_1,t_2,t_3)&=&\int\rmd E_1\rmd E_2\rmd
E_3\Bigl[\om_{\ve\hat n}(A^+_1A_3)\om_{\ve\hat
n}(A_1A^+_2)\om_{\ve\hat n}(A_2A^+_3)\\
&&+\om_{\ve\hat n}(A^+_1A_2)\om_{\ve\hat n}(A_1A^+_3)\om_{\ve\hat
n}(A^+_2A_3)\Bigr]
\end{eqnarray*}
This expression corresponds to the sum of the two
irreducible diagrams:

\noindent
\begin{picture}(120,10)
 \put(78,5){\line(1,0){75}} \put(78,5){\line(0,-1){13}} \put(153,5){\line(0,-1){13}}
 \put(94,0){\line(1,0){14}} \put(94,0){\line(0,-1){8}} \put(108,0){\line(0,-1){8}}
 \put(123,0){\line(1,0){14}} \put(123,0){\line(0,-1){8}} \put(137,0){\line(0,-1){8}}

 \put(181,5){\line(1,0){45}} \put(181,5){\line(0,-1){13}} \put(226,5){\line(0,-1){13}}
 \put(197,2){\line(1,0){43}} \put(197,2){\line(0,-1){10}} \put(240,2){\line(0,-1){10}}
 \put(211,-1){\line(1,0){45}} \put(211,-1){\line(0,-1){7}} \put(256,-1){\line(0,-1){7}}
\end{picture}
\[
A^+_1A_1A^+_2A_2A^+_3A_3+ A^+_1A_1A^+_2A_2A^+_3A_3
\]
In this case only the first diagram is non-zero in the
limit and Lemma~\ref{lemma} gives
\[
\fl\lim\limits_{\ve\to
0}W^T_\ve(t_1,t_2,t_3)=(2\pi)^2\dl(t_3-t_2)\dl(t_2-t_1)\int\rmd
E\langle g_3, P_E\hat nf_1\rangle\langle g_1,P_Ef_2\rangle\langle
g_2,P_Ef_3\rangle
\]
\end{example}
The case of arbitrary $n$ is described by the following
theorem.
\begin{theorem}\label{th1}
Let $T_l=|f_l\rangle\langle g_l|$, where $f_l,g_l\in{\cal
H}$ for $l=1,2,\dots,n$. One has the limit in the sense of
distributions in variables $t_1,\dots, t_n$:
\begin{eqnarray}
\lim\limits_{\ve\to 0}W^T_{\ve,\hat n,T_1,\om_1,\dots,
T_n,\om_n}(t_1,\dots,
t_n)=(2\pi)^{n-1}\dl(t_2-t_1)\dots\dl(t_n-t_{n-1})\nonumber\\
\times\dl_{\tilde\om_1,0} \int\rmd E\langle g_n,P_E\hat
nf_1\rangle\langle g_1,P_{E+\tilde\om_2}f_2\rangle\dots \langle
g_{n-1},P_{E+\tilde\om_n}f_n\rangle\label{t2}
\end{eqnarray}
For each $n$ only the following irreducible diagram is
non-zero as $\ep\to 0$:\\
\noindent
\begin{picture}(120,10)
 \put(78,6){\line(1,0){163}} \put(78,6){\line(0,-1){14}} \put(241,6){\line(0,-1){14}}
 \put(94,0){\line(1,0){14}} \put(94,0){\line(0,-1){8}} \put(108,0){\line(0,-1){8}}
 \put(123,0){\line(1,0){14}} \put(123,0){\line(0,-1){8}} \put(137,0){\line(0,-1){8}}
 \put(153,0){\line(1,0){14}} \put(153,0){\line(0,-1){8}} \put(167,0){\line(0,-1){8}}
 \put(200,0){\line(1,0){26}} \put(200,0){\line(0,-1){8}} \put(226,0){\line(0,-1){8}}
\end{picture}
\begin{equation}\label{diag}
A^+_1A_1A^+_2A_2A^+_3A_3A^+_4\dots A_{n-1}A^+_nA_n
\end{equation}
\end{theorem}
{\bf Proof.} Case (a): $\om_1=\om_2=\dots=\om_n=0$. Using
the correlation functions
\begin{eqnarray*}
\om_{\ve\hat n}(A^+_{i_\alpha}A_{j_\alpha})&=&\rme^{\rmi(t_{i_\alpha}-t_{j_\alpha})E_{i_\alpha}/\ve}\langle
g_{j_\alpha},\hat nf_{i_\alpha}\rangle\\
\om_{\ve\hat n}(A_{j_\beta}A^+_{i_\beta})&=&\frac{\rme^{\rmi(t_{i_\beta}-t_{j_\beta})E_{i_\beta}/\ve}
}{\ve}\langle g_{j_\beta},(1+\ve\hat n)f_{i_\beta}\rangle
\end{eqnarray*}
one can write~(\ref{10}) as
\begin{equation}\label{10.2}
\frac{1}{\ve^n}\exp\Bigl\{\rmi\Bigl[(t_1-t_{j_1})E_1+
\dots+(t_{i_n}-t_{j_n})E_{i_n}\Bigr]\Bigl/\ve\Bigr\}
\Bigl(\ve^kF(E)+O(\ve^{k+1})\Bigr)
\end{equation}
where
\[
F(E)=\prod\limits_{l=1}^k\langle g_{j_l},P_{E_l}
\hat nf_{i_l}\rangle \prod\limits_{l=k+1}^n\langle
g_{j_l},P_{E_{i_l}}f_{i_l}\rangle
\]

Define the permutations $p_i$ and $p_j$ of the set
$(1,\dots,n)$ by $p_i(l)=i_l$ and $p_j(l)=j_l$ for
$l=1,\dots,n$ and let $p_\alpha=p_ip^{-1}_j$. Consider the
expression in the square brackets in the exponent
in~(\ref{10.2}). The term proportional to $t_l$ in this
expression has the form $t_l(E_l-E_{\alpha_l})$, where
$\alpha_l=p_\alpha(l)$. Thus~(\ref{10.2}) can be written
as
\[
\frac{1}{\ve^n}\exp\Bigl\{\rmi\Bigl[t_n(E_n-E_{\alpha_n})+
\dots+t_1(E_1-E_{\alpha_1})\Bigr]\Bigl/\ve\Bigr\}\Bigl(\ve^kF(E)+O(\ve^{k+1})\Bigr)
\]
and with the notations
$\Om_l(E)=E_n+\dots+E_l-E_{\alpha_n}-\dots-E_{\alpha_l}$
for $l=2,\dots,n$ as
\begin{equation}
\frac{\rme^{\rmi(t_n-t_{n-1})\Om_n(E)/\ve}}{\ve}\dots
\frac{\rme^{\rmi(t_2-t_1)\Om_2(E)/\ve}}{\ve}
\Bigl(\ve^{k-1}F(E)+O(\ve^k)\Bigr)\label{eq9}
\end{equation}
If the expression~(\ref{10}) corresponds to an irreducible
diagram then the functions $\Om_l(E)$ are linearly
independent and, since they are linear in their arguments,
the convolution $\dl(\Om_2(E))\dots\dl(\Om_n(E))$ is well
defined.

In the case $k>1$, since for any $l=2,\dots,n$ (see
Lemma~\ref{lemma}):
\begin{equation}\label{limit}
\lim\limits_{\ve\to 0} \frac{\rme^{\rmi(t_l-t_{l-1})
\Om_l(E)/\ve}}{\ve} = 2\pi\dl(t_l-t_{l-1})\dl(\Om_l(E))
\end{equation}
and $k-1>0$, the limit of~(\ref{eq9}) equals to zero.

In the case $k=1$ the expression~(\ref{10}) corresponds to
the diagram~(\ref{diag}) and one has
\begin{eqnarray}&&
\om_{\ve\hat n}(A^+_1A_n)\om_{\ve\hat n}(A_1A^+_2)\dots
\om_{\ve\hat n}(A_{n-1}A^+_n)\nonumber\\
&=&\frac{\rme^{\rmi(t_n-t_{n-1})\Om_n(E)/\ve}}{\ve}\dots
\frac{\rme^{\rmi(t_2-t_1)\Om_2(E)/\ve}}{\ve}
\Bigl(F(E)+O(\ve)\Bigr)\label{eq12}
\end{eqnarray}
where $\Om_l(E)=E_l-E_1$. Using~(\ref{limit}) one finds
that the limit of the r.h.s. of~(\ref{eq12}) is
\begin{eqnarray*}
&&(2\pi)^{n-1}\dl(t_2-t_1)\dots\dl(t_n-t_{n-1})\dl(E_2-E_1)\dots\dl(E_n-E_1)\\
&&\times\langle g_n,P_{E_1}\hat nf_1\rangle
\langle g_1,P_{E_2}f_2\rangle\dots\langle g_{n-1},P_{E_n}f_n\rangle
\end{eqnarray*}
Integration over $E_1\dots E_n$ gives the
equality~(\ref{t2}) in the case (a).

Case (b): arbitrary $\om_1,\dots,\om_n$. In this case the
expression~(\ref{eq12}) in the decomposition~(\ref{eq1})
is multiplied by the factor $\exp(-\rmi\sum_l\om_l
t_l/\ep)$. The product can be written as
\[
\frac{\rme^{\rmi(t_n-t_{n-1})(\Omega_n(E)-\tilde\om_n)/\ve}}{\ve}\dots
\frac{\rme^{\rmi(t_2-t_1)(\Omega_2(E)-\tilde\om_2)/\ve}}{\ve}\rme^{-\rmi
t_1\tilde\om_1/\ve} \Bigl(F(E)+O(\ve)\Bigr)
\]
If $\tilde\om_1=0$ then the statement of the theorem
follows by the same arguments as in the case (a). If
$\tilde\om_1\ne 0$ then the limit of this term equals to
zero by Riemann-Lebesgue lemma due to the presence of the
rapidly oscillating factor $\exp(-\rmi
t_1\tilde\om_1/\ve)$. \hspace{1.2cm}$\Box$

\section{THE FREE WHITE NOISE NUMBER OPERATORS}
In the present section we show that the limiting truncated
correlation functions coincide with the complete (i.e.,
non-truncated) correlation functions of the free white
noise number operators.

\begin{definition}
{\bf Free white noise operators} $N_T(t)$ are the
operators satisfying the multiplication rule
\begin{equation}\label{eq6}
N_{T}(t)N_{T'}(t')=\dl(t-t')N_{T*T'}(t)
\end{equation}
where the $*$-product of any two operators $T$ and $T'$ is
defined by $T*T':=2\pi\int\rmd E P_ETP_ET'$.
\end{definition}
\begin{remark}
We call the operators $N_T(t)$ as {\bf free (or Boltzmann)
number operators} since they can be constructed using the
creation and annihilation operators $B^\pm_f(t)$
satisfying the free relations
$B^-_f(t)B^+_g(t')=2\pi\dl(t-t')\langle f,g\rangle$. In
fact, define $N_{|f\rangle\langle g|}(t):=\int\rmd E
B^+_{P_Ef}(t)B^-_{P_Eg}(t)$ and extend this definition by
linearity to any $T$. Then such defined operators satisfy
the relation~(\ref{eq6}).
\end{remark}

Let $\cal A$ be the algebra generated by the free white
noise operators $N_T(t)$ and let $\phi_{\hat n}$ be the
state on $\cal A$ characterized by $\phi_{\hat
n}(N_T(t))=\Tr(\hat nT)$.
\begin{theorem}
One has the equality
\begin{equation}\label{eq7}
\lim\limits_{\ep\to 0}W^{\rm T}_{\ep,\hat n,
T_1,0,\dots,T_n,0}(t_1,\dots,t_n)=\phi_{\hat
n}(N_{T_1}(t_1)\dots N_{T_n}(t_n))
\end{equation}
\end{theorem}
{\bf Proof.} By direct calculations using the
Eq.~(\ref{eq5}) and the relation~(\ref{eq6}).

The existence of the representation of the limiting
truncated correlation functions by the free white noise
number operators is related to the fact that only a
sub-class of the non-crossing irreducible diagrams
survives in the low density limit. We emphasize however,
that the l.h.s. of Eq.~(\ref{eq7}) is the limit of a
truncated correlation function whereas the r.h.s. contains
the complete correlation function.

\section{INDEPENDENCE AND THE GENERALIZED
POISSON STATISTICS IN THE LDL} The fact that the limiting
truncated correlation functions are the distributions in
variables $t_1,\dots,t_n$ with support at $t_1=\dots=t_n$
leads to the appearance of independent subalgebras in the
low density limit. In the beginning of this section we
remind the basic notions of independent subalgebras and of
cumulants. Then we find the asymptotically independent
subalgebras of ${\cal A}_\ep$ and discuss the limiting
statistics. We show that the cumulants and the moments of
certain elements in the algebra ${\cal A}_\ep$ in the low
density limit coincide with the cumulants and the moments
of the Poisson distribution.

\begin{definition}
Let $({\cal A},\om)$ be a $*$-probability space. A family
of unital $*$-subalgebras $\{{\cal A}_i\}_{i\in I}$,
${\cal A}_i\subset{\cal A}$, is called {\bf independent}
if $\om(a_1\dots a_n)=0$ whenever $a_l\in{\cal A}_{i_l}$,
$\om(a_l)=0$, and $k\ne l$ implies $i_k\ne i_l$.
\end{definition}

\begin{definition}
Let $({\cal A},\om)$ be a $*$-probability space. {\bf
Cumulants} of the space $({\cal A},\om)$ are the
multilinear functionals $\kappa_n\,:\,{\cal A}^n\to\mathbb
C,\, n\ge 1$, uniquely determined by
$\kappa_1(a):=\om(a),\, a\in{\cal A}$, and for $n>1$ by
induction through the relation:
\[
\om(a_1\dots
a_n)=\sum\limits_{\pi,\,\pi=:\{A_1,\dots,A_k\}}\kappa_{|A_j|}((a_1,\dots,a_n)|A_j)
\]
where the sum is over all partitions $\pi$ of the set
$\{1,\dots,n\}$ and "$(a_1,\dots,a_n)|A$" designates the
set of $a_i$ with $i\in A$.
\end{definition}

\begin{remark}\label{r1}
The cumulants $\kappa^{(\ep)}_n$ for a $*$-probability
space $({\cal A}_\ep,\om_{\ep\hat n})$ are directly
related to the truncated correlation functions. Namely, if
$a_1=N_{T_1,\om_1,\ep}(\varphi_1),\dots,
a_n=N_{T_n,\om_n,\ep}(\varphi_n)$, then
$\kappa^{(\ep)}_n(a_1,\dots,a_n)=W^T_{\ep,\hat
n,T_1,\om_1,\dots, T_n,\om_n}(\varphi_1,\dots,\varphi_n)$.
\end{remark}

For the analysis of independence in the low density limit
we introduce the notion of asymptotically independent
subalgebras for a $*$-probability space $({\cal
A}_\ep,\om_{\ep\hat n})$.

\begin{definition}
Let $({\cal A}_\ep,\om_{\ep\hat n})$ be a $*$-probability
space for the LDL. We say that a family of subalgebras
${\cal A}_{1,\ep},\dots,{\cal A}_{l,\ep}$ of ${\cal
A}_\ep$ is {\bf asymptotically independent} if
\[
\lim\limits_{\ep\to 0}\om_{\ep\hat n}(a_1,\dots,a_n)=0
\]
whenever $a_l\in{\cal A}_{i_l,\ep}$, $\om_{\ep\hat
n}(a_l)=0$, and $k\ne l$ implies $i_k\ne i_l$.
\end{definition}

The next theorem identifies asymptotically independent
subalgebras of ${\cal A}_\ep$.

\begin{theorem}\label{th2}
Let $\Lambda_1,\dots,\Lambda_l$ be a family of disjoint
open subsets in $\mathbb R$. Then the family of
subalgebras ${\cal A}_{\Lambda_1,\ep},\dots,{\cal
A}_{\Lambda_l,\ep}$ is asymptotically independent.
\end{theorem}
The proof follows from the fact that the truncated
correlation functions become in the limit as $\ep\to 0$
distributions in variables $t_1,\dots, t_n$ with support
at $t_1=t_2=\dots=t_n$.~$\Box$

Now let us analyze the statistics which appears in the low
density limit. From Theorem~\ref{mainth} and the relation
between the cumulants and the truncated correlation
functions it follows that $l$-th cumulant for the element
$a=N_{T,\om,\ep}(\varphi)$ in the limit has the form
\begin{equation}\label{eq4}\fl
\kappa_l(a,\dots,a)=\lim_{\ep\to 0}W^T_{\ep,\hat
n,T,\om,\dots,T,\om}(\varphi,\dots,\varphi)=\frac{1}{2\pi}\dl_{\om,0}
\int\rmd t\rmd E\Tr\hat n[2\pi\varphi(t) P_ET]^l
\end{equation}
We specify the further consideration to the case ${\cal
H}=L^2(\mathbb R^3)$. Consider $\hat n=1$ and
$S_t=\rme^{\rmi tH_1}$ where $H_1$ is the multiplication
operator by the function $\om(\bk)=|\bk|^2,\,\bk\in\mathbb
R^3$. Let $T_\lambda$ be an integral operator in $\cal H$
with the kernel
$T_\lambda(\bk,\bk')=(2\pi\sqrt{|\bk||\bk'|})^{-1}
\chi_{[0,\sqrt{\lambda}]}(|\bk|)\chi_{[0,\sqrt{\lambda}]}(|\bk'|)$,
where $\lambda$ is a positive number and
$\chi_{[0,\sqrt{\lambda}]}$ is the characteristic function
of the interval $[0,\sqrt{\lambda}]$. Let
$\varphi_0(t)=(2\pi)^{-1}\chi_{[0,2\pi]}(t)$.

\begin{theorem}\label{th4} Let $a_\lambda=N_{T_\lambda,\om,\ep}(\varphi_0)$, where
$T_\lambda$ and $\varphi_0$ are defined as above. Then for
any $l\in\mathbb N$ one has
\[
\kappa_l(a_\lambda,\dots,a_\lambda)=\lambda\dl_{\om,0}
\]
or equivalently, the cumulants of the element $a_\lambda$
with $\om=0$ coincide in the low density limit with the
cumulants of the Poisson distribution with expectation
equal to $\lambda$.
\end{theorem}
{\bf Proof.} The proof of the theorem is based on the
direct calculation of the cumulants using Eq.~(\ref{eq4}).
One has
\[
\frac{1}{2\pi}\int\rmd t[2\pi\varphi_0(t)]^l=1
\]
One also has
\begin{eqnarray*}
\fl\int\rmd E\Tr\hat n[P_ET_\lambda]^l&=&\int\rmd
E\int\rmd\bk_1\dots\rmd
\bk_l\dl(|\bk_1|^2-E)T_\lambda(\bk_1,\bk_2)\\
&&\times\dl(|\bk_2|^2-E)T_\lambda(\bk_2,\bk_3)\dots
\dl(|\bk_l|^2-E)T_\lambda(\bk_l,\bk_1)\\
&=&\int\rmd E\Bigl[T_\lambda\bigl(\sqrt{E},\sqrt{E}\bigr)
\int\rmd\bk\dl(|\bk|^2-E)\Bigr]^l=\int\rmd
E\chi_{[0,\sqrt{\lambda}]}\bigl(\sqrt{E}\bigr)=\lambda.
\end{eqnarray*}
Thus the r.h.s. of Eq.~(\ref{eq4}) equals to one. This
proves the theorem.\,\, $\Box$

Moments of the element $a_\lambda$ with $\om=0$ in the low
density limit are equal to the sum over all partitions of
the limiting cumulants and given by Touchard polynomials:
\[
\lim\limits_{\ep\to 0}\om_{\ep\hat
n}(a_\lambda^n)=\sum\limits_{k=1}^n S(n,k)\lambda^k
\]
where $S(n,k)$ is a Stirling number of the second kind,
i.e., the number of partitions of a set of size $n$ into
$k$ disjoint non-empty subsets. The limiting moments
coincide with the moments of the Poisson distribution with
expectation equal to $\lambda$. For $a_1$ one has
\[
\lim\limits_{\ep\to 0}\om_{\ep\hat n}(a_1^n)=B_n
\]
where $B_n$ is the $n$-th Bell number, i.e., the number of
partitions of a set of size $n$. The Bell numbers are the
moments of the Poisson distribution with expectation equal
to $1$.

\section{AN OPERATOR REPRESENTATION OF THE LIMITING
CORRELATION FUNCTIONS} In the present section we
explicitly realize the limiting correlation functions as
correlation functions of certain operators acting in a
suitable Hilbert space. Presence of delta functions in the
limiting correlation functions suggests that they can be
represented as correlation functions of certain white
noise operators. Here such a representation is constructed
in the special case using the results of\,\cite{apv1}.

Let $g_0, g_1\in{\cal H}$ satisfy the condition $\langle
g_0,S_tg_1\rangle=0$ for any $t\in\mathbb R$. Define for
$n,m=0,1$ the Hilbert space ${\cal K}_{nm}:=L^2({\rm
Spec}\, H_1,\rmd\mu_{nm})$, where ${\rm Spec}\,
H_1\subset\mathbb R$ is the spectrum of $H_1$ and
$\rmd\mu_{nm}:=\langle g_n,P_Eg_n\rangle\langle
g_m,P_E\hat ng_m\rangle\rmd E$. Let ${\cal
K}:=\bigoplus\limits_{n,m=0,1}{\cal K}_{nm}$ and let
${\cal H}_{WN}:=\Gamma(L^2(\mathbb R,{\cal K}))$ be the
symmetric Fock space over the Hilbert space of square
integrable $\cal K$-valued functions on $\mathbb R$
(abbreviation {\it WN} here stands for {\it White Noise}).
Using the natural decomposition ${\cal
H}_{WN}=\bigotimes\limits_{n,m=0,1}\Gamma(L^2(\mathbb
R,{\cal K}_{nm}))$ one can define the creation and
annihilation operator valued distributions
$B^\pm_{m,n}(E,t)$ acting in ${\cal H}_{WN}$ and
satisfying the canonical commutation relations:
\begin{equation}
\fl [B^-_{m,n}(E,t),B^+_{m',n'}(E',t')]=2\pi
 \dl(t'-t)\dl(E'-E)\langle g_m,P_Eg_{m'}\rangle
 \langle g_{n'},P_E\hat ng_n\rangle\label{cr1}
\end{equation}
The operator valued distributions $B^\pm_{m,n}(E,t)$ are
called {\bf time-energy quantum white noise} due to the
presence of $\dl(t'-t)\dl(E-E')$ in~(\ref{cr1}). Let
define the number operators
\[
\tilde N_{m,n}(E,t):=\sum\limits_{n'=0,1}\frac{1}{\langle
g_{n'},P_E\hat
ng_{n'}\rangle}B^+_{m,n'}(E,t)B^-_{n,n'}(E,t)
\]
and denote $N_{g_m,g_n}(t):=\int\rmd E[\tilde
N_{m,n}(E,t)+B^-_{n,m}(E,t)+B^+_{m,n}(E,t)]$. Let $\Omega\in{\cal
H}_{WN}$ be the vacuum vector.

\begin{theorem}\label{th3}
Let $T_1=|g_{m_1}\rangle\langle g_{n_1}|,\dots,
T_k=|g_{m_k}\rangle\langle g_{n_k}|$, where
$m_1,n_1,\dots,m_k,n_k\in\{0,1\}$. One has the equality
\begin{equation}\label{t1}
\lim\limits_{\ep\to 0}W_{\ep,\hat
n,T_1,0,\dots,T_k,0}(t_1,\dots,t_k)
=\langle\Omega,N_{g_{m_1},g_{n_1}}(t_1)\dots
N_{g_{m_k},g_{n_k}}(t_k)\Omega\rangle
\end{equation}
\end{theorem}
{\bf Proof.} r.h.s. of~(\ref{t1}) has the form
\begin{eqnarray*}\fl
\langle\Omega,N_{g_{m_1},g_{n_1}}(t_1)\dots
N_{g_{m_k},g_{n_k}}(t_k) \Omega\rangle =\int\rmd
E_1\dots\rmd E_k
\langle\Omega,[\tilde N_{m_1,n_1}(E_1,t_1)+B^-_{n_1,m_1}(E_1,t_1)\\
\fl +B^+_{m_1,n_1}(E_1,t_1)] \dots[\tilde N_{m_k,n_k}(E_k,t_k)
+B^-_{n_k,m_k}(E_k,t_k)+B^+_{m_k,n_k}(E_k,t_k)] \Omega\rangle
\end{eqnarray*}
Let us denote $\tilde N_{m,n}(t):=\int\rmd E \tilde
N_{m,n}(E,t)$. The truncated correlation function
corresponds to the term
\begin{equation}\label{te}\fl
\int\rmd E\rmd E' \langle\Omega,B^-_{n_1,m_1}(E,t_1)\tilde
N_{m_2,n_2}(t_2)\tilde N_{m_3,n_3}(t_3)\dots \tilde
N_{m_{k-1},n_{k-1}}(t_{k-1})B^+_{m_k,n_k}(E',t_k)\Omega\rangle
\end{equation}
Notice that $\tilde N_{m,n}(t)\Omega=0$. Therefore~(\ref{te})
equals to
\[\fl
\int\rmd E\rmd E'
\langle\Omega,[\dots[[B^-_{n_1,m_1}(E,t_1),\tilde
N_{m_2,n_2}(t_2)],\tilde N_{m_3,n_3}(t_3)]\dots \tilde
N_{m_{k-1},n_{k-1}}(t_{k-1})]
B^+_{m_k,n_k}(E',t_k)\Omega\rangle
\]
The commutators can be calculated by induction using the
canonical commutation relations~(\ref{cr1}). The result is
\begin{eqnarray}
\fl(2\pi)^{k-2}\dl(t_2-t_1)\dots\dl(t_{k-1}-t_{k-2})\int&&\rmd
E\rmd E'\langle g_{n_1},P_Eg_{m_2}\rangle\dots\langle
g_{n_{k-2}},P_Eg_{m_{k-1}}\rangle\nonumber\\
&&\times\langle\Omega,B^-_{n_{k-1},m_1}(E,t_{k-1})B^+_{m_k,n_k}(E',t_k)\Omega\rangle\label{1a}
\end{eqnarray}
The last two-point correlation function can be calculated using the
commutation relations~(\ref{cr1}). This gives for~(\ref{1a}) the
expression
\[\fl
(2\pi)^{k-1}\dl(t_2-t_1)\dots\dl(t_k-t_{k-1})\int\rmd E\langle g_{n_1},P_Eg_{m_2}\rangle\dots\langle
g_{n_{k-1}},P_Eg_{m_1}\rangle\langle g_{n_k},P_E\hat ng_{m_1} \rangle
\]
which coincides with the r.h.s. of~(\ref{t2}) in the case
$\om_1=\dots=\om_k=0$. \hspace{2.7cm}$\Box$

\begin{remark}
The limiting correlation functions could be represented as
expectations of certain quantum white noise operators in
the general case if one could construct a Hilbert space
${\cal H}_{WN}$, a vector $\Omega\in{\cal H}_{WN}$, and
operator valued distributions $B^\pm_{f,g}(E,t)$ and
$\tilde N_{f,g}(E,t)$ in ${\cal H}_{WN}$ with the property
$B^-_{f,g}(E,t)\Omega=\tilde N_{f,g}(E,t)\Omega=0$ and
satisfying the commutation relations
\begin{eqnarray}
{}[B^-_{f,g}(E,t),B^+_{f',g'}(E',t')]&=&2\pi\dl(t'-t)\dl(E'-E)\langle
f,P_Ef'\rangle \langle g',P_E\hat ng\rangle\label{cr11}\\
{}[B^-_{f,g}(E,t),\tilde
N_{f',g'}(E',t')]&=&2\pi\dl(t'-t)\dl(E-E')
\langle f,P_Ef'\rangle B^-_{g',g}(E,t)\label{cr12}\\
{}[\tilde N_{f,g}(E,t),\tilde
N_{f',g'}(E',t')]&=&2\pi\dl(t'-t)\dl(E'-E)[\langle g,P_Ef'\rangle N_{f,g'}(E,t)\nonumber\\
&&-\langle g',P_Ef\rangle N_{f',g}(E,t)]\label{cr13}
\end{eqnarray}
Suppose there exist such operators. Define
$N_{f,g}(t):=\int\rmd E[\tilde
N_{f,g}(E,t)+B^-_{g,f}(E,t)+B^+_{f,g}(E,t)]$. Then one can
prove exactly in the same way as in Theorem~\ref{th3} that
\[
\lim\limits_{\ep\to 0}W_{\ep,\hat n,|f_1\rangle\langle
g_1|,0,\dots, |f_n\rangle\langle g_n|,0}(t_1,\dots,
t_n)=\langle\Omega,N_{f_1,g_1}(t_1)\dots
N_{f_n,g_n}(t_n)\Omega\rangle
\]
\end{remark}

\section*{ACKNOWLEDGEMENTS} The author is grateful to
Luigi Accardi for useful discussions and for kind
hospitality in the Centro Vito Volterra of Rome University
"Tor Vergata" where a significant part of this work was
done. Special thanks to an anonymous referee for several
suggestions improving the quality and content of the
paper. The author acknowledges partial support from the
grant RFFI-05-01-00884-a.

\section*{APPENDIX A}
Here we make a connection between the objects defined in
section~II and the model of a test particle interacting
with a dilute Bose gas (see Ref.\cite{p} for details).

The one particle Hilbert space for this model has the form
${\cal H}\equiv L^2(\mathbb R^3)$, where $\mathbb R^3$ is
the 3-dimensional coordinate or momentum space. The one
particle free evolution is a unitary group
$S_t\equiv\rme^{\rmi t H_1}$ whose generator $H_1$ in the
momentum representation is the multiplication operator by
the function $\om(\bk)=|\bk|^2/2m$, where $m$ is the mass
of a gas particle. The test particle is characterized by
its Hilbert space ${\cal H}_{\rm S}$ and its free
Hamiltonian $H_{\rm S}$ acting in ${\cal H}_{\rm S}$ which
is assumed to have a discrete spectrum. The discrete set
$B$ is the set of all transition frequencies of the test
particle, or equivalently, the spectrum of its free
Liouvillean $-\rmi[H_{\rm S},\cdot]$.

The dynamics of a test particle interacting with a gas is
described by an evolution operator $U(t)$ acting in ${\cal
H}_{\rm S}\otimes\Gamma({\cal H})$ and satisfying in the
interaction picture, after the time rescaling $t\to
t/\ep$, the following Schr\"odinger equation
\begin{equation}\label{eq3}
\frac{\rmd U(t/\ep)}{\rmd t}=-\rmi\Bigl[\sum_{l,\om}
Q_{l,\om}\otimes N_{T_l,\om,\ep}(t)\Bigr] U(t/\ep)
\end{equation}
Here $Q_{l,\om}$ are certain operators in ${\cal H}_{\rm
S}$ such that $[H_{\rm S},Q_{l,\om}]=-\om Q_{l,\om}$ and
$T_l$ are certain operators in $\cal H$. The explicit form
of these operators is determined by the details of the
microscopic interaction between the test particle and
particles of the gas. Equation~(\ref{eq3}) is the place
where the operators $N_{T,\om,\ep}(t)$ appear.

The condition $S_{-t}\hat nS_t=\hat n$ and positivity of
$\hat n$ imply that for this model $\hat n$ is a
multiplication operator by a function $n:\mathbb R^3\to
[0,\infty)$. The value $n(\bk)$ has the meaning of the
density of gas particles at momentum $\bk$. If the state
of the gas is $\om_{\ep\hat n}$ then the density of gas
particles and the rate of collisions between the test
particle and the gas are of order $\ep$. Thus the limit
$\ep\to 0$ is the the low density limit. The limit is
non-trivial since the dynamics is studied on the kinetic
time scale of order $1/\ep$.

\section*{APPENDIX B}
Let $S(\mathbb R)$ be the Schwartz space over $\mathbb R$
and let $S'(\mathbb R)$ be the dual space of
distributions. We reproduce the following lemma
from\cite{alv}.
\begin{lemma}\label{lemma}
One has the limit in $S'(\mathbb R)\times S'(\mathbb R)$
\[
\lim\limits_{\ve\to 0}\frac{\rme^{\rmi
tx/\ve}}{\ve}=2\pi\dl(t)\dl(x)
\]
\end{lemma}
\noindent{\bf Proof.} Let $f,\phi\in S(\mathbb R)$ and let
$\tilde f$ be Fourier transform of $f$, $\tilde
f(\tau)=\int\rmd x\rme^{\rmi\tau x}f(x)$. One has the
identities
\begin{equation*}
 \fl I:=\lim\limits_{\ve\to 0}\int\rmd t\rmd x\frac{\rme^{\rmi
 tx/\ve}}{\ve}f(x)\phi(t)=
 \lim\limits_{\ve\to 0}\int\rmd\tau\phi(\ve\tau)\int\rmd x\rme^{\rmi\tau
 x}f(x)= \lim\limits_{\ve\to 0}\int\rmd\tau\phi(\ve\tau)\tilde f(\tau)
\end{equation*}
Since $\tilde f\in S(\mathbb R)$, the function
$\phi(\ve\tau)\tilde f(\tau)$ satisfies the conditions of
the Lebesgue lemma which allows to exchange the limit and
integration in the last expression. Thus
\[
I=\phi(0)\int\rmd\tau\tilde
f(\tau)=2\pi\phi(0)f(0)\hspace{6.9cm}\Box
\]

\end{document}